\documentclass{jetpl}

\twocolumn

\lat

\title{Internal friction as possible key factor governing the thermosensitivity of TRP channels}

\rtitle{Internal friction as possible key factor governing \ldots}

\sodtitle{Internal friction as possible key factor governing the thermosensitivity of TRP channels}

\author{A.\,O.\,Okenov$^+$,
	B.\,I.\,Iaparov$^+$, A.\,S.\,Moskvin$^+$\/\thanks{e-mail: alexander.moskvin@urfu.ru}}

\rauthor{A.\,O.\,Okenov, B.\,I.\,Iaparov, A.\,S.\,Moskvin}

\sodauthor{Okenov, Iaparov, Moskvin}

\address{$^+$Ural Federal University,
	620083 Ekaterinburg, Russia}

\abstract{Six thermo\d activated transient receptor potential (TRP) channels are the molecular basis of the thermosensation for mammals. But the molecular source of their gating remains unknown. In the Letter, we suggest a physically based model for the TRP channels and show that the temperature dependence of the "internal friction"\, can be a key factor governing the ion channels gating. Results of the computer modeling allowed us to successfully reproduce the experimental data for the open probability $P_{open}$ of the TRPV1 and TRPM8 channels at different temperatures and voltages.}

\PACS{74.50.+r, 74.80.Fp}

\begin{document}
	
	\maketitle
	
	\section{Introduction}
	
	The mammalian sensory system is capable of detecting and discriminating thermal stimuli over a broad temperature range from noxious cold ($\leq$\,8$^{\circ}$C) to noxious heat ($\geq$\,52$^{\circ}$C). Accumulated evidence suggests that the principal temperature sensors in the sensory nerve endings of mammals belong to the transient receptor potential (TRP) superfamily of cation channels~[1,\,2]. 	
	The temperature sensitivity factor $Q_{10}$ for the TRP channels can reach extremely large magnitudes up to 100~[3].
	
	What can be a reason for such an extraordinary temperature dependence is still an open question. The simplest model for any regulated ion channel is a two\d state model with one close (nonconductive) state and one open (conductive) state with the following Markov chain: $C \rightleftharpoons O$. The $P_o$ for this model is defined by the Boltzmann equation:
	\begin{equation}
	P_o=\frac{1}{1 + \exp \left(\frac{\Delta H - T\Delta S - zFV}{RT}\right)},
	\end{equation}
	where $\Delta H$ is the difference in enthalpy, $\Delta S$ the difference in entropy between the $O$ and $C$ state, $z$ the gating charge, $V$ the transmembrane voltage, $F$ the Faraday constant and $R$ the universal gas constant.
	
	Despite being really simple, the two\d state model can adequately describe the  experimental data on $P_o$ and the voltage value $V_{1/2}$ for the half\d maximal activation~[4] both for the  heat\d activated TRPV1 channel ($\geq 43^o\,C$) and for the cold\d activate TRPM8 channel (up to 26$^o\,C$), however, only with unphysically large parameters, e.g. the $\Delta H$ has to be in range $\Delta H = 44\ch150$\,kcal/mol ($1.9\ch 6.5$\,eV )~[4\ch6] for TRPV1.
	
	For this reason, Clapham and Miller~[7] suggested that the TRPs gating is accompanied by large changes in the molar heat capacity $\Delta C_p$ ($1.9$\ch$6.5$\,kJ\,mol$^{-1}$\,K$^{-1}$). According to the laws of thermodynamics, $\Delta C_p$ determines the change of $\Delta H(T)=\Delta H(T_0) + \Delta C_p(T-T_0)$ and $\Delta S(T)=\Delta S(T_0) + \Delta C_p \ln(T/T_0)$, where~$T_0$ is an arbitrarily chosen reference temperature. However, the hypothesis leads to an unexpected conclusion: every thermoTRP has to be both heat\d and cold\d activated.  Such symmetry is not confirmed experimentally.
	
	We should note that the both approaches overlook the effect of the strong temperature sensitivity of the intra\d channel  conformational dynamics due to a strong temperature dependence of so\d called internal friction~[8]. The effect was addressed recently for ryanodine receptors (RyR)~[9] which reveal a marked temperature dependence of the open probability~[10]. The use of the electron\d conformational (EC) model~[11] and the assumption of the Arrhenius temperature dependence of  internal friction allowed to explain all the temperature effects in the ligand\d gated RyRs.
	
	In the paper we apply  a similar approach for the explanation of the TRP channels gating and show that the EC\d model given the assumption of the Arrhenius temperature dependence of internal friction does provide an adequate description of the steep temperature dependence $P_o(T)$ with the physically reasonable values of the energy parameters.
	
	\section{Internal friction}

	Popular concept of the potential energy landscape in a multidimensional configurational (conformational) space with multiple (meta)stable conformational states~[12] presents one of starting points for the description of the protein dynamics. However, because of limited knowledge about the protein structure and the incomprehensible complexity of the energy landscape we have to tolerate considerable simplifications in the theoretical approaches,  especially since determining local properties of the energy landscape by experiment is very challenging.
	
	In reality, a channel can pass through numerous conformations during gating, all of which are in a continuum. However, for practical purposes, only those states that represent the lowest points in the energy profile can be considered discrete states stable enough to be deduced from single\d channel data. Within different Markovian kinetic models of the channel  gating these states fall into three categories, that is different closed states, different open states, and so\d called inactivation states.
	
	Quite different approach  to describe the essential protein dynamics with the multidimensional potential energy landscape implies a projection from the full configuration space to low dimensionality. Projection onto a low\d dimensional energy surface with effective configurational (conformational) coordinate (collective variable) puts protein transformations on the same footing as other chemical reactions in solution, which can be described by simple Kramers theory that implies effective energy landscape as a function of a reaction coordinate as an abstract one\d dimensional coordinate which represents progress along a reaction pathway.
	
	Dimensionality reduction effects inherent in the projection generally slow the dynamics when the reaction coordinate does not account for barriers in the potential energy landscape corresponding to full dimensionality. The effect can be taken into account by introduction of the effective  ''internal friction'' with a phenomenological friction parameter $\Gamma$~[8]. This effect is sometimes associated with ''roughness'', manifested as a position\d dependent effective diffusion coefficient, which affects the physical interpretation of this phenomenological parameter. The original barriers on the potential energy surface are therefore reflected in two different ways in a low\d dimensional projection. The barriers along the reaction coordinate are still explicit, but barriers in ''orthogonal'' degrees of freedom correspond to position dependent diffusion coefficients. Some groups describe both the explicit and implicit barriers as ''roughness''~[12].
	
	According to Rauscher~[8], the external reaction coordinate that represents the protein’s displacement regarding the environment\t is characterized by a flat energy landscape, while the effective parameter~$\Gamma$ compensates the additional roughness.

	Thus, the internal friction is an effective ''intra\d protein'', parameter that characterizes interactions between protein atoms, reflects a roughness of the energy landscape, accumulates the effects of the projection procedure, and depends on temperature~[8] and reaction coordinates~[13]. Its temperature dependence can be approximated by a simple Arrhenius law:
	\begin{equation}
	\Gamma=\Gamma_0\exp\left(E_\Gamma /k_BT \right),
	\end{equation}
	where $E_{\Gamma}$ is the internal friction activation energy, $\Gamma_0$ is a temperature independent parameter.
	
	\section{Electron\d conformational model}
	
	The simplest one\d dimensional model of the channel implies existence of the low\d energy conformational potential branch with the two (meta)stable states conditionally termed as ''closed'' and ''open'', respectively, separated by an energy barrier. Such a branch can be most simply deduced in the framework of a simple ''two\d level'' electron\d conformational (EC) model~[11]. 	
	The simplest (''toy'') EC-model starts with the two ''electronic'' states whose energy can strongly depend on the conformational coordinate $Q$. The electronic states can be associated with the ligand binding/unbinding. Modeling the ion channels we emphasize a specific role of the breathing mode $Q$ and start with a simple and a little bit naive picture of the massive nanoscopic channel like an elastic rubber tube with a varying cross-section governed by a conformational breathing coordinate $Q$.
	
	As a starting point of the ''toy'' EC\d model algebra we introduce a simple effective Hamiltonian for a single ion channel  with the two states as follows:
	\begin{equation}
	H_s= - \Delta\hat{\sigma}_z-h\hat{\sigma}_x - pQ + \frac{KQ^2}{2} + aQ\hat{\sigma}_z,
	\label{Hs}
	\end{equation}
	where $\hat{\sigma}_x$ and 	$\hat{\sigma}_z$ are the Pauli matrices, and the first term describes the bare (i.e. at $Q = 0$) energy splitting of ''up'' and ''down'' states. The second term describes a quantum ''mixing'' effect.
	The third, linear in $Q$ term in~(\ref{Hs}) corresponds to the energy of an external conformational stress, described by an effective stress, or pressure parameter $p$. The fourth term in~(\ref{Hs}) implies a simple harmonic approximation for the conformational energy, $K$ being the effective ''elastic'' constant.
	The last term in~(\ref{Hs}) describes a linear EC interaction with the coupling parameter $a$. Conformational variable $Q$ is the dimensionless one, therefore the effective parameters ($h$, $p$, $K$, $a$) are assigned the energy units. It is worth noting that the Hamiltonian (\ref{Hs}) resembles that of a  two-level electronic system interacting with a bosonic mode~[14].
	
	Two eigenvalues of the Hamiltonian:
	\begin{equation}
	E_{\pm}(Q)=\frac{KQ^2}{2} - pQ \pm \frac{1}{2}\sqrt{(\Delta - aQ)^2 + h^2}
	\label{E}	
	\end{equation}
	define the two branches of the conformational potential, attributed to electronically closed ($E_-$) and electronically open ($E_+$) states of the channel. Given $h = 0$ we arrive at two diabatic potentials $E_\pm(Q)$ for electronically closed and open states, shown in Fig.\,1 at $\Delta = 0$. Two minima are separated by an intersection point at $Q = 0$ thus providing the ''bistability'' conditions. It should be noted that for $h \neq 0$ instead of the two parabolas we arrive at two novel isolated EC branches where instead of the twofold degenerated intersection point we have a high\d energy metastable state and low\d energy unstable transition state (see inset in Fig.\,1).
	
	The energy gap $\Delta (Q=0)$ obviously depends on the membrane potential $V$. Taking into account the linear and quadratic terms, this dependence can be represented as follows
	\begin{equation}
	\Delta = \Delta_0 + k_1 V + \frac{1}{2}k_2 V^2.
	\end{equation}
	where $\Delta_0$ is the energy difference at zero membrane potential, $k_1$ is a measure of the effective total charge that interacts with the membrane potential, $k_2$ describes the dielectric properties of the channel and surrounding lipids~[15,\,16]. Since positive values of $\Delta$ stabilize the closed state and negative values stabilize the open state, $k_1$ must be negative in case of activation by depolarization and vice versa. It should be noted that the stress parameter $p$ which  describes the mechanosensitivity of the channel also depends on $V$, however, below we take into account only  the voltage dependence of $\Delta$.
	
	\begin{figure}[t]
		\centering
		\includegraphics[width=82mm, natwidth=82mm, natheight=75mm]{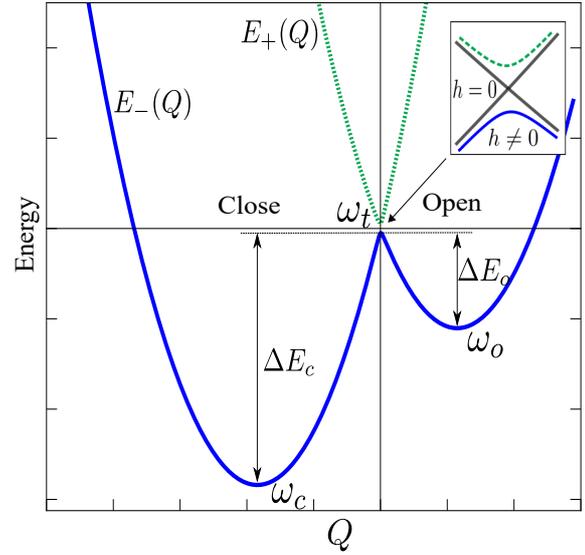}
		\caption{{\bf Fig.\,1.} The energy profile of the EC-model for the TRP channel with a global minimum for the closed state. The subscripts ''c'' and ''o''  correspond to the close and open states, $\Delta E_c$, $\Delta E_o$ are potential barriers heights,  $\omega_o$, $\omega_c$ are angular frequencies inside (meta)stable minima and $\omega_t$ is the angular frequency at the transition state. Inset shows quantum effect of nonzero parameter $h$ pointing to formation of unconventional metastable state and unstable transition state}
		\label{fig:ecm}
	\end{figure}
	
	The conformational dynamics of the channel is assumed to obey a conventional Langevin equation of motion:
	\begin{equation}
	M\ddot {Q}= - M\frac{\partial}{\partial Q}E(Q) - M\Gamma\dot Q + \xi(t)\,,
	\label{L}
	\end{equation}
	where $M$ is an effective mass (below M let to be unity), $\Gamma$ is an internal friction, $\xi(t)$ is the Gaussian\d Markovian noise:$\langle \xi (t) \rangle = 0\,; \,\,
	\langle \xi (t)\xi (t') \rangle = 2\gamma k_B T M \delta (t-t')$, where $\gamma$ is an ''external'' friction parameter.
	
	As we are modeling a nonequilibirum, open, multicomponent, and multimodal system, the dissipation  of conformational  dynamics is separated  into ''internal'' and ''external'' ones with different values of the friction parameters. The external dissipation occurs due to interactions between the surface atoms of an ion channel and the intracellular and extracellular solutions; $\gamma$ varies directly with the viscosity $\eta$ of the solvent. The  internal dissipation occurs due to the effective intra\d protein viscosity. Unfortunately, there is no reliable experimental data even about the viscosity of the cytoplasm and about its temperature dependence. But, in practice, both $\Gamma$ and $\gamma$ are assumed to obey the Arrhenius dependence on temperature~[8]:
	\begin{equation}
	\gamma=\gamma_0\exp\left(E_\gamma/k_BT \right).
	\end{equation}
	
	Hereafter, we neglect the temperature dependence of the main EC\d model parameters and suppose that $\Gamma$'s are constant within the potential well:
	\begin{equation}
	\Gamma (Q)=
	\begin{cases}
	\Gamma_o = \Gamma_o^0\exp\left({E^o_\Gamma}/{k_B T} \right), & Q \geq Q_t\\
	\Gamma_c = \Gamma_c^0\exp\left( {E^c_\Gamma}/{k_B T} \right), & Q < Q_t
	\end{cases}
	,
	\end{equation}
	where $Q_t$ separates ''open'' and ''close'' states; it is determined by a coordinate of the transition state.
	
	Obviously, the EC\d model can be applied for description of different types of the ionic channels from mechanosensitive and voltage\d gated channels to the ligand\d activated ones. Indeed, the stress parameter $p$ regulates the effect of external pressure; the $\Delta$ parameter depends on the membrane potential. The ligand activation implies transition between different conformational branches.
	
	\section{Effective temperature and Kramers reaction rate theory}
	Hereafter we'll address  the low\d energy double\d walled branch $E_-$ of the conformational potential to be a starting point to solve the Kramers problem, that is to find the rate at which a Brownian particle escapes from a metastable state over a potential barrier~[17]. The main difference is that we have two different frictions (''external'' and ''internal'') instead of one. In this section we show how to reduce our model to the Kramers escape theory.
	
	First, we rewrite the Langevin equation~(13) as system of the first order stochastic differential equations and determine the drift ($D_v$, $D_Q$) and diffusion ($D_{QQ}$, $D_{Qv}$, $D_{vv}$ ) coefficients.
	\begin{equation}
	\frac{dQ}{dt} = v;
	\frac{dv}{dt} = f(Q) - \Gamma v + \xi(t),
	\label{LL}
	\end{equation}
	where $v$ is a velocity and the force $f(Q) = - \frac{\partial}{\partial Q} E(Q)$.
	
	Second, we insert results to the general Fokker\d Planck equation for the probability distribution function $\rho(Q, v, t)$~[18]:
	\begin{equation}
	\frac{\partial \rho}{\partial t} = \left[-\frac{\partial}{\partial Q} v - \frac{\partial}{\partial v} [-\Gamma v + f(Q)] + \gamma k_B T \frac{{\partial}^2}{\partial v^2}\right]\rho.
	\end{equation}
	The Fokker\d Plank equation has a stationary solution $\rho(Q,v) = \frac{\Gamma}{2 \gamma k_B T} e^{ \left[ - \frac{\Gamma}{\gamma k_B T}\left(\frac{v^2}{2} + E(Q)\right) \right]}$, which can be considered as an effective Boltzmann distribution at the effective temperature $T^* = \frac{\gamma}{\Gamma}T$ (see~[19]).
	
	Third, we replace absolute temperature ($T$) by the effective temperature ($T^*$) in the Langevin equation~(\ref{L}), and obtain a new correlation function, $\xi (t)$:
	\begin{align}
	\langle \xi (t) \rangle = 0\,; \,\,
	\langle \xi (t)\xi (t') \rangle = 2 \Gamma k_B T^* \delta (t-t')
	\label{xi}
	\end{align}
	
	Now, for the Langevin equation~(\ref{L}) with the correlation function~(\ref{xi}), we can directly use the Kramers results for escape rates. For a strong friction limit, the Kramers theory gives a particularly simple expression
	\begin{equation}
	k=\frac{A}{\gamma}\exp\left( - E_a/k_BT \right) \, ,
	\label{k}
	\end{equation}
	where $A$ is a temperature and viscosity independent constant. The expression is most commonly used in protein folding dynamics. But, to be accurate at high temperatures, we use more general equation for the escape rate, which corresponds to a moderate\d to\d large friction limit~[17]:
	\begin{equation}
	k_{c} = \left(\sqrt{\frac{\Gamma_c^2}{4} + \omega_t^2} - \frac{\Gamma_c}{2}\right)\frac{\omega_c}{2\pi\omega_t}\exp \left(-\frac{\Gamma_c}{\gamma}\frac{\Delta E_c}{k_B T} \right) \, ,
	\label{k_c}
	\end{equation}
	where $k_c$ is an escape rate from ''close'' state (Fig.\,1). Angular frequency inside metastable minimum, $\omega_c$, and angular frequency at the transition state, $\omega_t$, are determined as:
	\begin{equation}
	\omega_c = \sqrt{\partial^2 E/\partial Q^2\big\rvert_{Q_c}}\,; \,\,
	\omega_t = \sqrt{|\partial^2 E/\partial Q^2|\big\rvert_{Q_t}}\,.
	\end{equation}
	The equation for moderate\d to\d strong friction limit~(\ref{k_c}) reduces to the ''classic'' Kramers equation~(\ref{k}) if strong friction limit is satisfied ($\Gamma \gg \omega_t$) and if the Langevin dynamics obeys the fluctuation\d dissipation theorem ($\Gamma \sim \gamma$).
	
	The validity limit of the Kramers escape rate~(20) is determined by the two relevant dimensionless parameters: the inverse van\,'t\,Hoff\d Arrhenius factor $k_B T^*/\Delta E$ and the friction strength $\Gamma/\omega_t$. As internal friction parameters strongly depend on temperature, equation~(\ref{k_c}) might be invalid at low and high temperatures.
	
	The probability of the channel transition to the open state is determined as follows:
	\begin{equation}
	P_{o} = 1/(1 + {k_{o}}/{k_{c}}).
	\end{equation}
	where $k_{o}$ is an escape rate from open state that came similarly with~(\ref{k_c}).
	
	\begin{table*}[t]
		\label{tab:ecmparam}
		\caption{{\bf Tab.\,1.} EC\d model parameters}
		\centering
		\begin{tabular} {ccccccccccc}
			\hline
			& $K$ & $a$ & $h$ & $p$ & $E_\Gamma^c$ & $E_\Gamma^o$ & $E_{\gamma}$ & $\Gamma_c^0$ & $\Gamma_o^0$ & $\gamma_0$ \\ \hline
			TRPV1 & 16.7 & 45.0 & 1.0 & -6.8 & 890 & 75 & 660 & $9.80 \cdot 10^{-14}$ & 4.89 & $2.53 \cdot 10^{-10}$ \\
			TRPM8 & 15.7 & 40.0 & 1.0 & -6.8 & 123 & 670 & 490 & 0.924 & $3.8 \cdot 10^{-10}$ & $1.32 \cdot 10^{-7}$ \\  \hline
		\end{tabular}
	\end{table*}
	
	\begin{figure*}[t]
		\centering
		\includegraphics[width=160mm, natwidth=160mm, natheight=75mm]{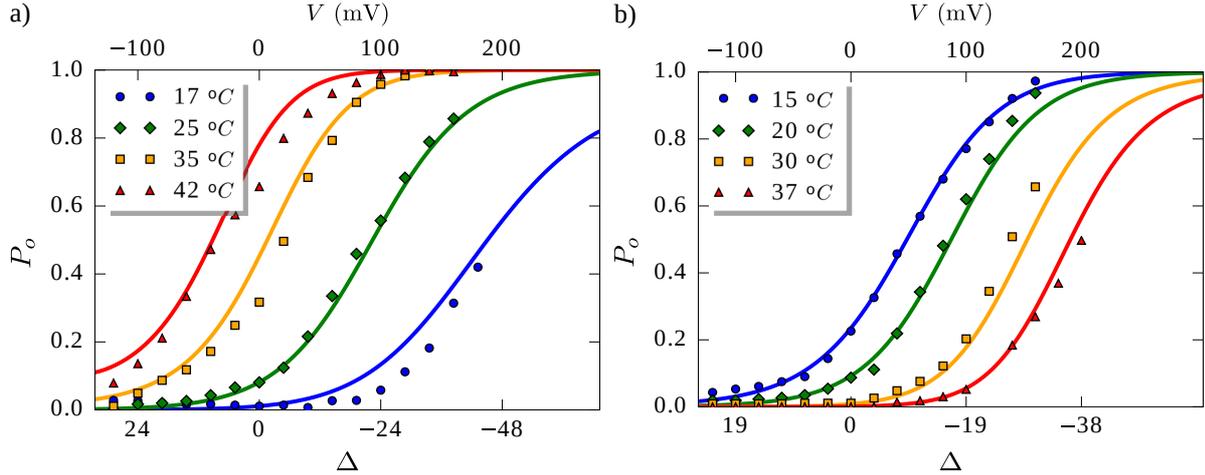}
		\caption{{\bf Fig.\,2.} Voltage dependence of steady\d state activation curves at different temperatures for: a) TRPV1 channel; b) TRPM8 channel. Solid lines represent the EC model prediction. Experimental values are adapted from~[4].}
		\label{fig:pod}
	\end{figure*}
	
	To confirm the Kramers results by a numerical solution, we implemented second order integrator~[20] for the Langevin equations~(\ref{LL}) in C++ programming language. We obtained activation curves at different temperatures and calculated the half\d maximal activation voltage.
	
	\section{Results and discussion}
	
	''Working'' parameters of the EC\d model for the TRP channels are unknown, so we will rather arbitrarily choose them based on a physically reasonable assumption about the height of barriers $\Delta E_c$ and $\Delta E_o$ of the order of 1\,kcal/mol, or 43.3\,meV (Tab.\,1). As a requirement for the conformational potential to remain double\d welled, the parameter $\Delta$ must be in a range (-70.8\ch34.1)\,meV for TRPV1 and (-61.1\ch26.4)\,meV for TRPM8. These sets of parameters ensure that barrier height will be less than 52.4\,meV in case of TRPV1 and 43.8\,meV in case of TRPM8. We selected the values of the friction parameters (activation energies and pre\d exponential factors) (Tab.\,1) to fit experimental data from Ref.~[4] for the both channels. All the parameters, except $\Gamma_c^0$, $\Gamma_o^0$, and $\gamma_0$, are presented in meV.
	
	Assuming a simple linear dependence $\Delta (V)$, we obtained the voltage\d dependent steady\d state activation curves for both channels at different temperatures (Fig.\,2) with $\Delta_0=0$ and $k_1=-0.24$ for TRPV1 and $\Delta_0=0$ and $k_1=-0.19$ for TRPM8. The voltage does fully open and close the channel only in a limited temperature range and the slope of the activation curves changes with temperature as predicted by the allosteric model. In contrast, the two\d state model and MWC (Monod\d Wyman\d Changeux) model predict that voltage can fully open and close channels, regardless the temperature~[3]; temperature gradually shifts the voltage\d activation curves without altering  the slope.	
	
	Voltage becomes a partial activator at high and low temperatures. For TRPV1 the thermal energy $k_BT^*_c$ is enough to escape from ''close'' state at high $T$ for any $\Delta$ while at low $T$, the thermal energy $k_BT^*_o$, which decreases with $T$ for $E_\Gamma^o<E_{\gamma}$, is enough to escape from ''open'' state for any $\Delta$. The effect is opposite for TRPM8: at high $T$ the thermal energy $k_BT^*_o$ is enough to escape from ''open'' state; at low $T$ the thermal energy $k_BT^*_c$ is enough to escape from ''close'' state.
	
	\begin{figure*}[t]
		\label{fig:ponum}
		\centering
		\includegraphics[width=160mm, natwidth=160mm, natheight=75mm]{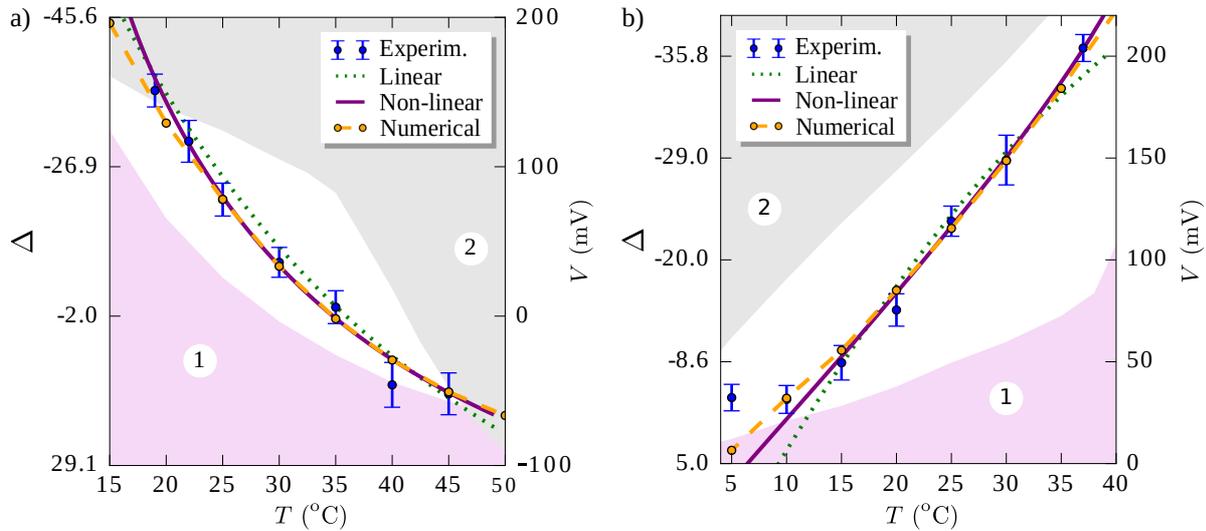}
		\caption{{\bf Fig.\,3.} $V_{1/2}$ as a function of temperature for: a) TRPV1 channel; b) TRPM8 channel. The dotted line corresponds to the linear $\Delta (V)$ dependence; the solid line does to the nonlinear one; the dashed line with dots represents results of numerical solution of Langevin equation. (Note, left axis is unrelated to the linear dependence)}
	\end{figure*}
	
	In the Kramers approximation, the analytical expression for the half\d maximal activation voltage $V_{1/2}$ can be presented only in implicit form; the final expression is complex and uninformative. Assuming the linear dependence $\Delta (V)$, we calculated $V_{1/2}(T)$ curves (Fig.\,3). The curves diverge from the prediction of the two\d state model especially at hyperpolarization and high depolarization. Adding quadratic term in $\Delta (V)$ ($\Delta_0=-2; k_1=-0.28; k_2=0.62\cdot10^{-3}$ for TRPV1 and $\Delta_0=5; k_1=-0.295; k_2=0.91\cdot10^{-3}$ for TRPM8) leads to increasing the curvature of $V_{1/2}$ (Fig.\,3) or even changing the curvature to the opposite one (Fig.\,3b). The nonlinear $\Delta (V)$ dependence gives a better fit for the experimental data. By filling in Fig.\,3 we have identified the  areas where  the Kramers equation for escape rate is invalid for open (1) and close (2) states, respectively. Accordingly, numerical calculations show that the $V_{1/2}(T)$  curves deviate noticeably from the predictions of the Kramers theory at low temperatures.

	\section{Conclusion}
	
	Making use of the physically based EC model for the thermo\d activated TRP channels we have shown that the temperature dependence of the internal friction can be a key factor governing the ion channels gating with much smaller energy barriers $\leq$\,0.1\,eV.  EC model was simplified to a double well potential with dynamics which is analytically described in terms of the Kramers rate theory. Results of the computer modeling allowed us to successfully reproduce the experimental voltage dependence of the open probability $P_{open}$ and the  temperature dependence of the half\d maximal activation voltage $V_{1/2}$. We argue that the second\d order term in $\Delta (V)$ is necessary to accurately describe the TRP channels gating given such a wide range of temperatures. The internal friction mechanism is believed to open novel perspectives for studying the high temperature sensitivity of the TRP channels and other channels as well.
	
	Supported by Act 211 Government of the Russian Federation, agreement No. 02.A03.21.0006 and by the Ministry of Education and Science, projects No. 2277 and No. 5719.

\end{document}